\documentclass[12pt]{article}

\usepackage{scicite}

\input psfig.sty

\newenvironment{sciabstract}{%
\begin{quote} \bf}
{\end{quote}}

\newcounter{lastnote}

\title{An Asymmetric, Energetic Type Ic Supernova \\
Viewed Off-Axis, and a Link to Gamma-Ray Bursts
}

\author
{Paolo~A.~Mazzali$^{1,2,3,4\ast}$,
Koji S.~Kawabata$^{5}$,
Keiichi~Maeda$^{6}$, Ken'ichi~Nomoto$^{1,2\ast\ast}$,\\
Alexei V. Filippenko$^{7}$,
Enrico Ramirez-Ruiz$^{8}$,
Stefano Benetti$^{9}$,
Elena~Pian$^{4}$,\\
Jinsong~Deng$^{1,2,10}$,
Nozomu~Tominaga$^{1}$,
Youichi~Ohyama$^{11,12}$,
Masanori~Iye$^{1,13,14}$,\\
Ryan J. Foley$^{7}$,
Thomas Matheson$^{15}$,
Lifan~Wang$^{16}$, and
Avishay Gal-Yam$^{17}$\\
\\
\footnotesize{$^{1}$Department of Astronomy, School of Science,}
\footnotesize{University of Tokyo, Bunkyo-ku, Tokyo 113-0033, Japan}\\
\footnotesize{$^{2}$Research Center for the Early Universe, School of Science,}
\footnotesize{University of Tokyo, Bunkyo-ku, Tokyo 113-0033, Japan}\\
\footnotesize{$^{3}$Max-Planck Institut f\"ur Astrophysik,}
\footnotesize{Karl-Schwarzschild Str. 1, D-85748 Garching, Germany}\\
\footnotesize{$^{4}$Istituto Nazionale di Astrofisica-OATs, Via Tiepolo 11,
I-34131 Trieste, Italy}\\ 
\footnotesize{$^{5}$Hiroshima Astrophysical Science Center, 
Hiroshima University, Hiroshima 739-8526, Japan}\\ 
\footnotesize{$^{6}$Department of Earth Science and Astronomy,
College of Arts and Sciences, }\\
\footnotesize{University of Tokyo,Komaba 3-8-1, Meguro-ku, Tokyo 153-8902, Japan}\\ 
\footnotesize{$^{7}$Department of Astronomy, University of 
California, Berkeley, CA 94720-3411, USA}\\ 
\footnotesize{$^{8}$Institute for Advanced Study, Einstein Drive, Princeton, 
NJ 08540, USA; Chandra Fellow}\\ 
\footnotesize{$^{9}$INAF-Osservatorio Astronomico di Padova, 
vicolo del'Osservatorio 5, 35122 Padova, Italy}\\ 
\footnotesize{$^{10}$National Astronomical Observatories, 
CAS 20A Datun Road,}
\footnotesize{Chaoyang District, Beijing 100012, China}\\
\footnotesize{$^{11}$Department of Infrared Astrophysics,
Institute of Space and Astronautical Science, }\\
\footnotesize{Japan Aerospace Exploration Agency, 
3-1-1 Yoshinodai, Sagamihara, Kanagawa, 229-8510, Japan}\\
\footnotesize{$^{12}$Subaru Telescope, NAOJ, 650 North A'ohoku Place, 
Hilo, HI 96720, USA}\\ 
\footnotesize{$^{13}$Optical and IR Astronomy Division,
National Astronomical Observatory,}
\footnotesize{Mitaka, Tokyo 181-8588, Japan}\\ 
\footnotesize{$^{14}$Department of Astronomical Science, 
School of Physical Sciences,}\\
\footnotesize{Graduate University for Advanced Studies,
2-21-1 Osawa, Mitaka, Tokyo 181-8588, Japan}\\ 
\footnotesize{$^{15}$National Optical Astronomy Observatory, 950
N. Cherry Avenue, Tucson,}
\footnotesize{AZ 85719-4933, USA}\\ 
\footnotesize{$^{16}$Lawrence Berkeley National Laboratory, 50-232, 1 
Cyclotron Road, Berkeley, CA 94720, USA}\\ 
\footnotesize{$^{17}$Department of Astronomy, California Institute of
Technology, }
\footnotesize{Pasadena, CA 91125, USA; Hubble Fellow}\\ 
\footnotesize{$^\ast$mazzali@ts.astro.it, $^{\ast\ast}$nomoto@astron.s.u-tokyo.ac.jp}\\
\normalsize{{\bf To be published in Science on 27 May 2005.}}
}

\date{}

\newcommand{\Msun}{M_{\odot}}
\newcommand{\OI}{O~{\sc i}}
\newcommand{\MgI}{Mg~{\sc i}}
\newcommand{\NaI}{Na~{\sc i}}
\newcommand{\CaII}{Ca~{\sc ii}}
\newcommand{\FeII}{Fe~{\sc ii}}
\newcommand{\Fefs}{$^{56}$Fe}
\newcommand{\Cofs}{$^{56}$Co}
\newcommand{\Nifs}{$^{56}$Ni}

\def\gsim{\mathrel{\rlap{\lower 4pt \hbox{\hskip 1pt $\sim$}}\raise 1pt \hbox {$>$}}} 
\def\lsim{\mathrel{\rlap{\lower 4pt \hbox{\hskip 1pt $\sim$}}\raise 1pt \hbox {$<$}}}

\begin{document} 


\baselineskip24pt


\maketitle


\begin{sciabstract}
Type Ic supernovae, the explosions following the core collapse of massive stars 
that have previously lost their hydrogen and helium envelopes, 
are particularly interesting because of the link with long-duration gamma-ray bursts. 
Although indications exist that these explosions are aspherical, direct evidence has 
still been missing. Late-time observations of SN~2003jd, a luminous Type Ic supernova, 
provide such evidence.  Recent Subaru and Keck spectra reveal double-peaked profiles 
in the nebular lines of neutral oxygen and magnesium.  
These profiles are different from those of known Type Ic supernovae, 
with or without a gamma-ray burst, and they can be understood if SN~2003jd was an aspherical, 
axisymmetric explosion viewed from near the equatorial plane. 
If SN~2003jd was associated with a gamma-ray burst, we missed the burst 
as it was pointing away from us.
\end{sciabstract}


When a massive star reaches the end of its life and exhausts its
nuclear fuel, the core itself collapses to form a compact remnant (a
neutron star or a black hole).  Although the exact mechanism is not
well-understood, the resulting release of energy leads to the ejection
of the envelope of the star at high velocities, producing a supernova
(SN).

Typically, a massive star has a large H-rich envelope, making it
difficult to observe the innermost part, where the action takes place.
There are, however, some cases where the H envelope, and also the
inner He envelope, were lost before the star exploded, through either
a stellar wind or, more likely, binary interaction ({\it 1}).  These SNe,
called Type Ic, offer a view close to the core, and so they are
particularly interesting as tools to study the properties of the
collapse and of the SN ejection.

Adding to this, some SNe~Ic, characterized by a very high kinetic
energy ({\it 2,3}), have been observed to be linked with the previously
unexplained phenomenon of gamma-ray bursts (GRBs) --- brief but
extremely bright flashes of hard ($\gamma$-ray and X-ray) radiation
which for decades had baffled astronomers ({\it 4--8}). The link between
SNe~Ic and GRBs is probably not accidental. If a jet is produced by a
collapsing star, it can only emerge and generate a GRB if the stellar
envelope does not interfere with it ({\it 9}).

In view of this link, we have searched among known SNe~Ic for the
counterpart of a property that is typical of GRBs: asphericity. A
jet-like explosion is required for GRBs from energetics
considerations: if they were spherically symmetric, GRBs would involve
excessively large energies, comparable to the rest mass of several
suns. The best evidence for asphericity in the GRB-associated SNe
(GRB/SNe) has so far come from the fact that iron seems to move faster
than oxygen in the ejected material. Evidence of this is seen in
spectra obtained several months after the explosion, when the ejected
material has decreased in density and behaves like a nebula. The
GRB/SN 1998bw ({\it 10}) showed strong emission lines of [\OI] (a forbidden
line of neutral oxygen), as do normal SNe~Ic, but also of [\FeII] (a
forbidden line of singly-ionized iron), which are weak in normal
SNe~Ic. The [\FeII] lines near 5100~\AA\ in SN 1998bw are broader than
the [\OI] 6300, 6363~\AA\ blend.

Asphericity can explain this peculiar situation ({\it 11}). In a typical,
spherical SN explosion, heavier elements are produced in deeper layers
of the progenitor star, and as a consequence of the hydrodynamical
properties of the explosion they are given less kinetic energy per
unit mass than external layers, which typically contain lighter
elements.  However, in a jet-like explosion the heavier elements (in
particular $^{56}$Ni) are probably synthesized near the jet at the
time of core collapse, and are ejected at high velocities. Lighter
elements such as oxygen, which are not produced in the explosion but
rather by the progenitor star during its evolution, are ejected near
the equatorial plane with a smaller kinetic energy, and are
distributed in a disc-like structure.

Given this scenario, the observed line profiles depend on the
orientation of the explosion with respect to our line of sight. Iron
can be observed to be approaching us at a higher velocity than oxygen
if we view the explosion near the jet direction, which is also the
requirement for the GRB to be observed ({\it 10,11}).  The [\OI] line, on
the other hand, will appear as a narrow, sharp line in the case of a
polar view, since in that case oxygen moves almost perpendicular
with respect to our line of sight (the case of SN~1998bw), but it will
show a broader, double-peaked profile for an equatorial view, since a
large fraction of the oxygen would then be moving either toward or
away from the observer ({\it 11}).

While this picture seems well established for GRB/SNe, it is important
to determine whether it may be common to other SNe~Ic. Measurements of
the relative widths of the Fe and O lines are difficult for fainter
SNe~Ic because the Fe lines are weak. However, the [\OI] line is
always rather strong, and it can be expected that, given a
sufficiently large sample, variations should be seen in its profile
reflecting different viewing angles.

So far the evidence for this was missing ({\it 12}), but recent observations
of SN~2003jd seem to close this gap. SN~2003jd, discovered on 25
Oct. 2003 ({\it 13}) (UT dates are used in this paper), 
is a SN~Ic at a distance of $\sim 80$\,Mpc and reached
a rather bright maximum. Assuming a Galactic extinction $E(B-V)_{\rm
Gal} = 0.06$ mag ({\it 14}), and a host extinction $E(B-V)_{\rm Host} =
0.09$ mag, as derived from the strength of the interstellar \NaI~D
absorption ({\it 15}), we derive $M_B({\rm Max}) \approx -18.7$ mag.  This
is much more luminous than the normal SN~Ic 1994I ({\it 1,16}), and is
comparable to the GRB/SNe 1998bw ({\it 4}) and 2003dh ({\it 6}).
Spectroscopically, however, SN~2003jd shows narrower lines than the
hyper-energetic GRB/SNe (Fig.\ 1), and it appears to be intermediate
between those and the normal SN~Ic 1994I ({\it 16}). The closest analogue
may be the energetic SN~2002ap ({\it 17,18}).

We observed SN~2003jd in the nebular phase with the Japanese 8.2-m
Subaru telescope on 12 Sep. 2004 ({\it 19}) 
using the FOCAS spectrograph ({\it 20}), and with the 10-m Keck-I
telescope on 19 Oct. 2004 using LRIS ({\it 21}). These dates correspond to
SN ages of $\sim$330 and $\sim$370 days after explosion,
respectively.  In both spectra (Fig.\ 2) the nebular line [\OI] 6300,
6363~\AA\ clearly has a double-peaked profile with FWHM~$\approx
8000$\,km\,s$^{-1}$. The \MgI] 4570~\AA\ line shows a similar
profile.  Magnesium is formed near oxygen in the progenitor star.  The
[\FeII] blend near 5100~\AA\ is quite weak.

Late-phase emission is created by the release of the heat deposited by
the $\gamma$-rays and the positrons emitted in the decay chain \Nifs\
$\rightarrow$ \Cofs\ $\rightarrow$ \Fefs. Therefore, the mass of
\Nifs\ can be determined indirectly through the strength of the
emission lines. Since SN~2003jd was not as luminous as SN~1998bw, we
rescaled the synthetic spectra to the appropriate \Nifs\ mass, the
best value for which was $\sim$0.3~$\Msun$. This is actually very
similar to that derived for the GRB-associated SN~2003dh ({\it 22,23}), and
much larger than in the non-GRB SNe~Ic (e.g., {\it 17}).

We computed nebular spectra of 2D explosion models for various
asphericities and orientations ({\it 11}).  We found that a spherical model
produces a flat-topped [\OI] profile, which is not compatible with either
SN 1998bw or SN 2003jd.  The flat-topped emission is a typical characteristic
of emission from a shell.  Indeed, the emission from any spherically
distributing materials should have the maximum at the wavelength of
the line transition between 6300--6363~\AA, taking into account the
line blending.  On the other hand, Fig.\ 3 shows that a highly
aspherical model can explain the [\OI] line profiles in both SN
1998bw and SN 2003jd.  To reproduce the double-peaked profile of the
[\OI] line in SN~2003jd we found that SN~2003jd must be oriented
$\gsim 70^\circ$ away from our line of sight. In contrast, for
SN~1998bw this angle was only $\sim 15$--$30^\circ$ ({\it 11}), and it was
even smaller for SN~2003dh ({\it 24}).  Less aspherical models do not
produce sufficiently sharp [\OI] in SN 1998bw.

This result confirms that SN 2003jd is a significantly aspherical
explosion, and raises the interesting question of whether SN~2003jd was
itself a GRB/SN. A GRB was not detected in coincidence with SN~2003jd
({\it 25}).

If the explosion was very off-axis, we do not anticipate to have been
able to detect $\gamma$-rays. However, a GRB is expected to produce a
long-lived radiative output through synchrotron emission. X-ray and
radio emission are produced by the deceleration of the relativistic
jet as it expands into the wind emitted by the progenitor star before
it exploded. This afterglow emission is very weak until the Doppler
cone of the beam intersects our line of sight, making off-axis GRB
jets directly detectable only months after the event, and at long
wavelengths.

SN~2003jd was observed in X-rays with Chandra on 10 Nov. 2004, about
30 days after the explosion, and was not detected to a limit $L_X \leq
3.8 \times 10^{38}$\,erg\,s$^{-1}$ in the energy interval 0.3--2 keV
({\it 26}).  It was also observed in the radio (8.4 GHz) 9 days after the
explosion, and again not detected to a limit $L_R \leq
10^{27}$\,erg\,s$^{-1}$\,Hz$^{-1}$ ({\it 27}).

These non-detections may suggest that SN~2003jd did not produce a
GRB. However, absence of evidence is not necessarily evidence of
absence.  Let us consider the standard jet associated with typical
GRBs, i.e., a sharp-edged, uniform jet with $E = 10^{51}$\,erg and a
$5^{\circ}$ opening angle, expanding laterally at the local sound
speed ({\it 28, 29}; see Fig.4 legend for other parameters).  If the jet
expands in a wind with density $\dot{M}/v = 5 \times
10^{11}$\,g\,cm$^{-1}$ (e.g., mass-loss rate $\dot{M} \sim
10^{-6}\Msun$\,yr$^{-1}$, velocity $v \sim 2000$\,km\,s$^{-1}$), it would
give rise to the X-ray and radio light curves shown in Fig.\ 4.  If
the SN made a large angle ($\geq 60^{\circ}$) with respect to our
line of sight, its associated GRB jet, if present, 
would not have been detected in the X-rays or radio.
Furthermore, the afterglow emission can be fainter for a
lower jet energy and/or a lower wind density.

While this does not by itself prove that SN~2003jd produced a GRB,
it is certainly a possibility, since quantities such as the ejected
mass of \Nifs\ are comparable to those typical of GRB/SNe
({\it 23}). Moreover, our observations confirm that the energetic SN
2003jd is an aspherical explosion, reinforcing the case for a link
with a GRB.  It could also be the case for other energetic SNe~Ic.
Note that the lack of X-ray and radio emission does not place
stringent constraints on the intrinsic kinetic energy carried by the
SN as long as the ejecta experience little deceleration before
$\sim$30 days. The expansion velocity (which need not be isotropic)
must be $\lsim 0.2 A_*^{-1/3} (\epsilon_e/0.1)^{-1/3} c$ (where
$\epsilon_e$ is the fraction of the total blast energy that goes into
shock-accelerated electrons) in order to produce $L_X (t\approx
30\;{\rm days}) \leq 3.8 \times 10^{38}$\,erg\,s$^{-1}$ for a standard
$10^{51}$ erg SN shell expanding into a wind with density
$A_*=(\dot{M}/v)/(5 \times 10^{11}$\,g\,cm$^{-1}) = 1$ ({\it 28}). The
average expansion velocity along our line of sight can be estimated
from the early-time spectra; for SN~2003jd this is about $0.05c$. In
an aspherical explosion, however, the kinetic energy must be
considerably larger near the rotation axis of the stellar
progenitor, with bulk expansion velocities close to $\sim 0.1c$.

The fact that the bright SN~2003jd is the first SN~Ic showing double
peaks in the [\OI] line ({\it 12}) suggests that the degree of asphericity
is not the same in all SNe~Ic.  The GRB/SNe (1998bw, 2003dh), which
are probably highly aspherical, have been discovered thanks to a GRB
trigger; their orientation is therefore such that the [\OI] profile
must be single-peaked.  For normal SNe~Ic, which are on average closer
and easier to discover, the lack of observed double-peaked profiles
suggests that they are not as strongly aspherical.  SN~2003jd appears
to share many of the properties (energetics, luminosity) of the
GRB/SNe, but it was discovered independent of a GRB, and it is likely
to be an aspherical SN viewed off-axis.  There have been three
energetic SNe~Ic without a GRB trigger (therefore less biased) --- SNe
1997dq, 1997ef, and 2002ap --- whose nebular spectra did not show the
double peaks in the [\OI] 6300~\AA.  Given the small sample, the number
(one out of four) is not inconsistent with our interpretation that the
viewing angle $\gsim 70^{\rm o}$ results in the double-peaked [\OI].

Additional sensitive radio and X-ray observations of SN 2003jd are
strongly encouraged, because a jet with the standard parameters could
be still detectable by deep observations (Fig. 4).  The result should
provide valuable tests for the presence of an off-axis GRB with the
typical parameters and would further constrain the viewing angle and the
mass-loss rate.
 

{\bf References and Notes}

\begin{enumerate}

\item
K. Nomoto {\it et al.,} 
{\it Nature} {\bf 371}, 227 (1994).
 
\item 
K. Iwamoto {\it et al.,} 
{\it Nature} {\bf 395}, 672 (1998).

\item 
K. Nomoto {\it et al.,} 
{\it Stellar Collapse}, ed. C. L. Fryer (Dordrecht: Kluwer), 277
(2004).

\item 
T.~J. Galama {\it et al.,} 
{\it Nature} {\bf 395}, 670 (1998).

\item 
C. Stanek {\it et al.,} 
{\it Astrophys. J.} {\bf 591}, L17 (2003).

\item 
T. Matheson {\it et al.,} 
{\it Astrophys. J.} {\bf 599}, 394 (2003).

\item 
J. Hjorth {\it et al.,} 
{\it Nature} {\bf 423}, 847 (2003).

\item 
J. Malesani {\it et al.,} 
{\it Astrophys. J.} {\bf 609}, L5 (2004).

\item 
A.E. MacFadyen, S.E. Woosley 
{\it Astrophys. J.} {\bf 524}, 262 (1999).
 
\item 
P.A. Mazzali {\it et al.,} 
{\it Astrophys. J.} {\bf 559}, 1047 (2001).

\item 
K. Maeda {\it et al.,} 
{\it Astrophys. J.} {\bf 565}, 405 (2002).

\item 
T. Matheson {\it et al.,} 
{\it Astron. J.} {\bf 121}, 1648 (2001).

\item 
J. Burket, B. Swift, W. Li 
{\it IAUC} {\bf 8232} (2003).

\item 
D. Schlegel, D. Finkbeiner, M. Davis 
{\it Astrophys. J.} {\bf 500}, 525 (1998). 
 
\item
S. Benetti {\it et al.,} in preparation.

\item 
A.V. Filippenko {\it et al.,} 
{\it Astrophys. J.} {\bf 450}, L11 (1995).

\item 
P.A. Mazzali {\it et al.,} 
{\it Astrophys. J.} {\bf 572}, L61 (2002).

\item 
R.J. Foley {\it et al.,} 
{\it Publ. Astron. Soc. Pacific} {\bf 115}, 1220 (2003).

\item 
K. Kawabata {\it et al.,} 
{\it IAU Circ.} {\bf 8410} (2004).

\item 
N. Kashikawa {\it et al.,} 
{\it Publ. Astron. Soc. Japan} {\bf 54}, 819 (2002).

\item 
J.B. Oke {\it et al.,} 
{\it Publ. Astron. Soc. Pacific} {\bf 107}, 375 (1995).

\item
J. Deng, N. Tominaga, P.A. Mazzali, K. Maeda, K. Nomoto 
{\it Astrophys. J.} {\bf 624}, 898 (2005) 

\item 
P.A. Mazzali {\it et al.,} 
{\it Astrophys. J.} {\bf 599}, L95 (2003).

\item 
E. Ramirez-Ruiz {\it et al.,} 
in preparation.

\item 
K. Hurley {\it et al.,} 
{\it GCN} {\bf 2439} (2003).

\item 
D. Watson, E. Pian, J.N. Reeves, J. Hjorth, K. Pedersen 
{\it GCN} {\bf 2445} (2003).

\item 
A.M. Soderberg, S.R. Kulkarni, D.A. Frail 
{\it GCN 2435} (2003).

\item 
E. Ramirez-Ruiz, P. Madau 
{\it Astrophys. J.} {\bf 608}, L89 (2004). 

\item 
J. Granot, E. Ramirez-Ruiz 
{\it Astrophys. J.} {\bf 609}, L9 (2004). 

\item 
F. Patat {\it et al.,} 
{\it Astrophys. J.} {\bf 555}, 900 (2001).

\item 
This work is based on data collected at the Subaru telescope
operated by NAOJ, and at the Keck telescopes made possible by the
W. M. Keck Foundation.  This research is supported in part by JSPS and
MEXT in Japan, and by NSF in the USA.

\end{enumerate}

%



\clearpage
\begin{figure}
\psfig{file=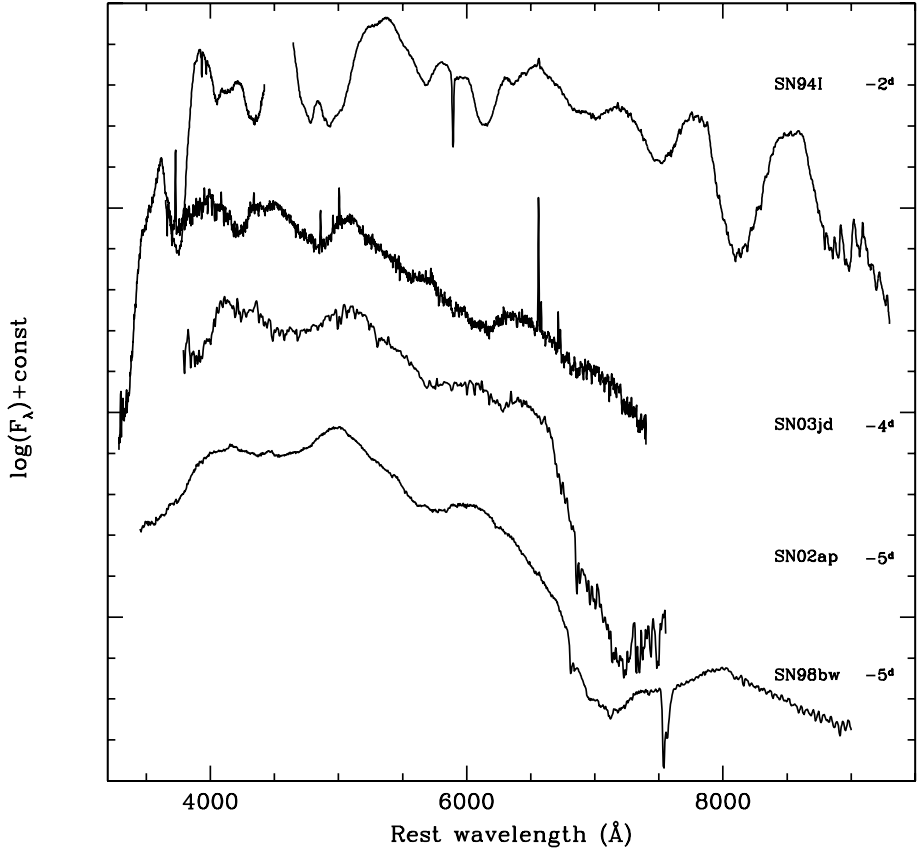,width=0.8\textwidth}
\caption[]{The near-maximum optical spectrum of SN~2003jd
compared with spectra of other SNe~Ic at a similar phase
($F_{\lambda}$ is the flux per unit wavelength).  The dates are the
days relative to the optical maximum (i.e., the minus sign means
before the maximum light).  Spectra are ordered by increasing line
width (implying increasing kinetic energy per unit mass), ranging from
the normal SN~1994I (16), to the energetic SN~2002ap (17,18), and to
the hyper-energetic GRB/SN 1998bw (30). The absorption line near
7600~\AA\ in the spectrum of SN 1998bw is telluric.
\label{fig1}}
\end{figure}

\clearpage
\begin{figure}
\psfig{file=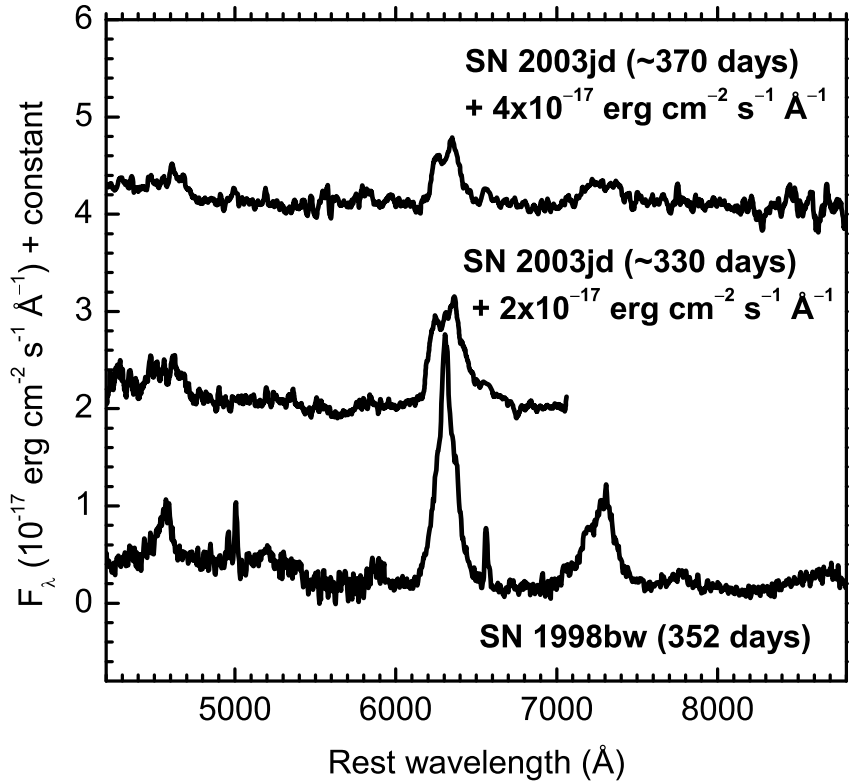,width=0.8\textwidth}
\caption[]{Nebular spectra of SNe~Ic.  Bottom: nebular
spectrum of SN~1998bw (30) taken 337 days after maximum light (352
days after the explosion).  Notice the \MgI], [\FeII], [\OI], and
[\CaII] lines near 4570, 5100, 6300, and 7300~\AA, respectively.
Middle: Subaru+FOCAS spectrum of SN~2003jd, $\sim$330 days after the
putative time of explosion.  Top: Keck spectrum of SN~2003jd at an
epoch of $\sim$370 days.  The [\OI] 6300, 6363~\AA\ line in SN~2003jd
clearly exhibits a double-peaked profile. Marginal evidence of a
double peak is also present in the profiles of \MgI] 4570~\AA\ and
[\CaII] 7300~\AA.  The spectrum of SN~1998bw has been shifted in flux
to make it consistent with the distance of SN~2003jd.
\label{fig2}}
\end{figure}

\clearpage
\begin{figure}
\psfig{file=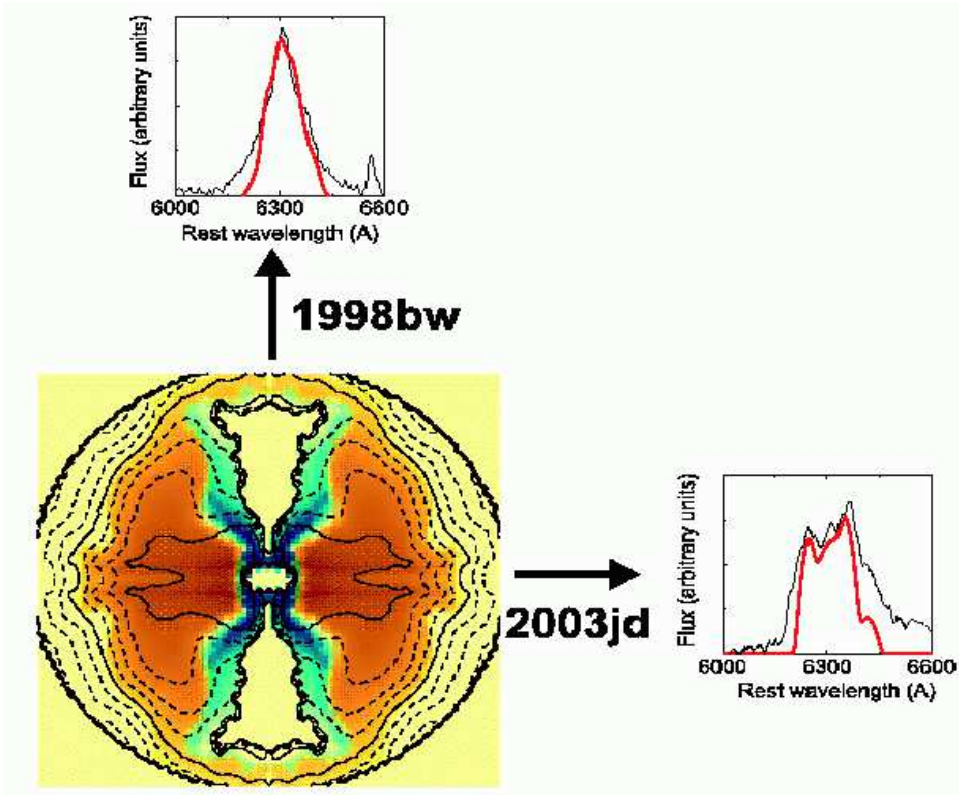,width=0.8\textwidth}
\caption[]{Nebular line profiles observed from an aspherical
explosion model depend on the orientation. The figure shows the
properties of the explosion model computed in 2D (11): Fe (colored in
blue) is ejected near the jet direction and oxygen (brown) in a
disc-like structure on and near the equatorial plane. Density contours
(covering 2 orders of magnitude and divided into 10 equal intervals in
log scale) reflect the dense disc-like structure.  Synthetic [\OI]
6300, 6363~\AA\ lines (red lines) computed in 2D are compared with
the spectra of SN~1998bw and SN~2003jd (black lines).
\label{fig3}}
\end{figure}

\clearpage
\begin{figure}
\psfig{file=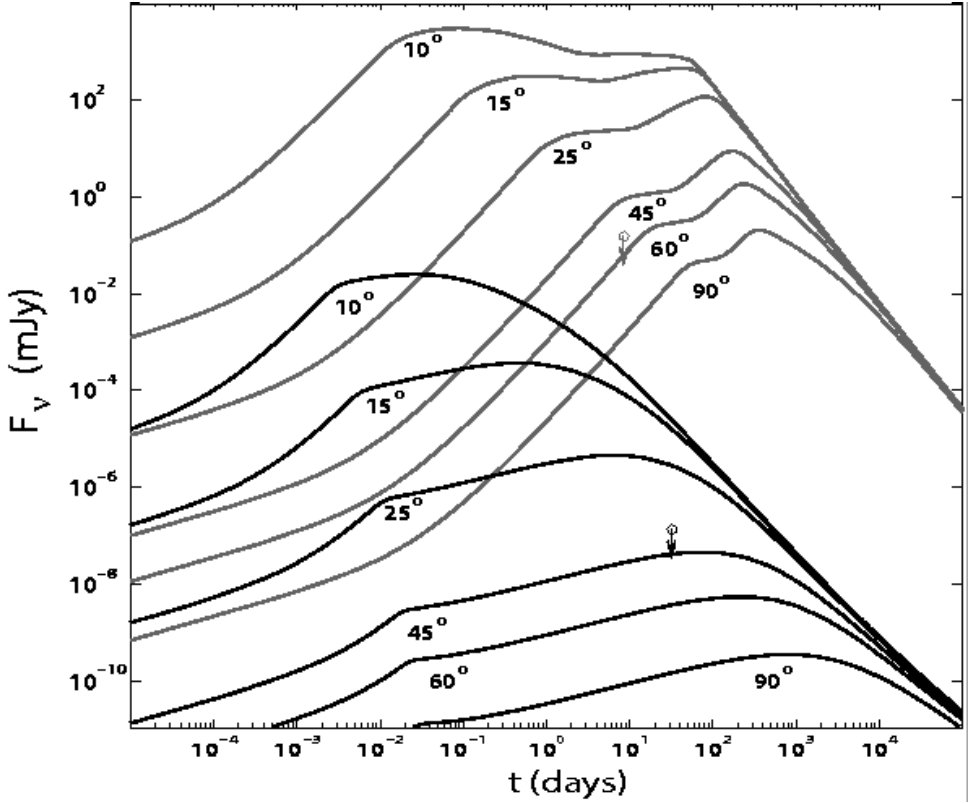,width=0.8\textwidth}
\caption[]{Afterglow emission from a sharp-edged, uniform jet
in SN 2003jd. X-ray (0.3--2 keV, black) and radio (8.4 GHz, gray) 
light curves are calculated for various viewing angles $\theta_{\rm obs}$
for a GRB with the standard parameters $E_{\rm jet}=10^{51}$ erg,
$\epsilon_e = 0.1$, $\epsilon_B = 0.1$, $\theta_0 = 5^\circ$, and
$A_*=1$ (where $E_{\rm jet}$ is the energy in the jet, $\epsilon_e$
and $\epsilon_B$ are the fraction of the internal energy in the
electrons and magnetic field, respectively, and $\theta_0$ is the
opening half-angle of the jet).  The synchrotron spectrum is taken to
be a piecewise power law with the usual self-absorption, cooling, and
injection frequencies calculated from the cooled electron distribution
and magnetic field (28, 29). The observed radio and X-ray upper limits
for SN~2003jd are marked by open circles.  Cosmological parameters
taken in the model are $\Omega_{\rm m} = 0.27$, $\Omega_{\Lambda} =
0.73$, and $H_0 = 72$ km s$^{-1}$ Mpc$^{-1}$.
\label{fig4}}
\end{figure}

\end{document}